\documentclass[sigconf]{acmart}
\usepackage{xcolor}
\usepackage{amsmath}
\usepackage{multirow}
\usepackage{enumitem}

\AtBeginDocument{%
  }

\copyrightyear{2024}
\acmYear{2024}
\setcopyright{rightsretained}
\acmConference[CHI EA '24]{Extended Abstracts of the CHI Conference on
Human Factors in Computing Systems}{May 11--16, 2024}{Honolulu, HI, USA}
\acmBooktitle{Extended Abstracts of the CHI Conference on Human Factors in
Computing Systems (CHI EA '24), May 11--16, 2024, Honolulu, HI, USA}
\acmDOI{10.1145/3613905.3651059}
\acmISBN{979-8-4007-0331-7/24/05}

\begin{document}

\title{Adaptive 3D UI Placement in Mixed Reality Using Deep Reinforcement Learning}

% \author{Anonymous for Review}
% \affiliation{%
%   \institution{JPMorgan Chase \& Co.}
%   \state{NY}
\author{Feiyu Lu}
\authornote{Both authors contributed equally to this research.}
\email{feiyu.lu@jpmchase.com}
\orcid{0000-0002-1939-9352}
\affiliation{%
  \institution{JPMorgan Chase \& Co.}
  \city{New York}
  \state{NY}
  \country{USA}
}
\author{Mengyu Chen}
\authornotemark[1]
\email{mengyu.chen@jpmchase.com}
\orcid{0000-0001-6833-7273}
\affiliation{%
  \institution{JPMorgan Chase \& Co.}
  \city{New York}
  \state{NY}
  \country{USA}
}
\author{Hsiang Hsu}
\email{hsiang.hsu@jpmchase.com}
\orcid{0000-0001-8084-3929}
\affiliation{%
  \institution{JPMorgan Chase \& Co.}
  \city{New York}
  \state{NY}
  \country{USA}
}
\author{Pranav Deshpande}
\email{pranav.deshpande@jpmchase.com}
\orcid{0009-0007-1612-9430}
\affiliation{%
  \institution{JPMorgan Chase \& Co.}
  \city{New York}
  \state{NY}
  \country{USA}
}
\author{Cheng Yao Wang}
\email{chengyao.wang@jpmchase.com}
\orcid{0000-0003-3229-4341}
\affiliation{%
  \institution{JPMorgan Chase \& Co.}
  \city{New York}
  \state{NY}
  \country{USA}
}
\author{Blair MacIntyre}
\email{blair.macintyre@jpmchase.com}
\orcid{0000-0002-5357-2366}
\affiliation{%
  \institution{JPMorgan Chase \& Co.}
  \city{New York}
  \state{NY}
  \country{USA}
}

\renewcommand{\shortauthors}{Feiyu Lu*, Mengyu Chen*, Hsiang Hsu, Pranav Deshpande, Cheng Yao Wang, and Blair MacIntyre}

\begin{abstract}
  Mixed Reality (MR) could assist users' tasks by continuously integrating virtual content with their view of the physical environment. However, where and how to place these content to best support the users has been a challenging problem due to the dynamic nature of MR experiences. In contrast to prior work that investigates optimization-based methods, we are exploring how reinforcement learning (RL) could assist with continuous 3D content placement that is aware of users' poses and their surrounding environments. Through an initial exploration and preliminary evaluation, our results demonstrate the potential of RL to position content that maximizes the reward for users on the go. We further identify future directions for research that could harness the power of RL for personalized and optimized UI and content placement in MR.
\end{abstract}

%%

\begin{teaserfigure}
  \includegraphics[width=\textwidth]{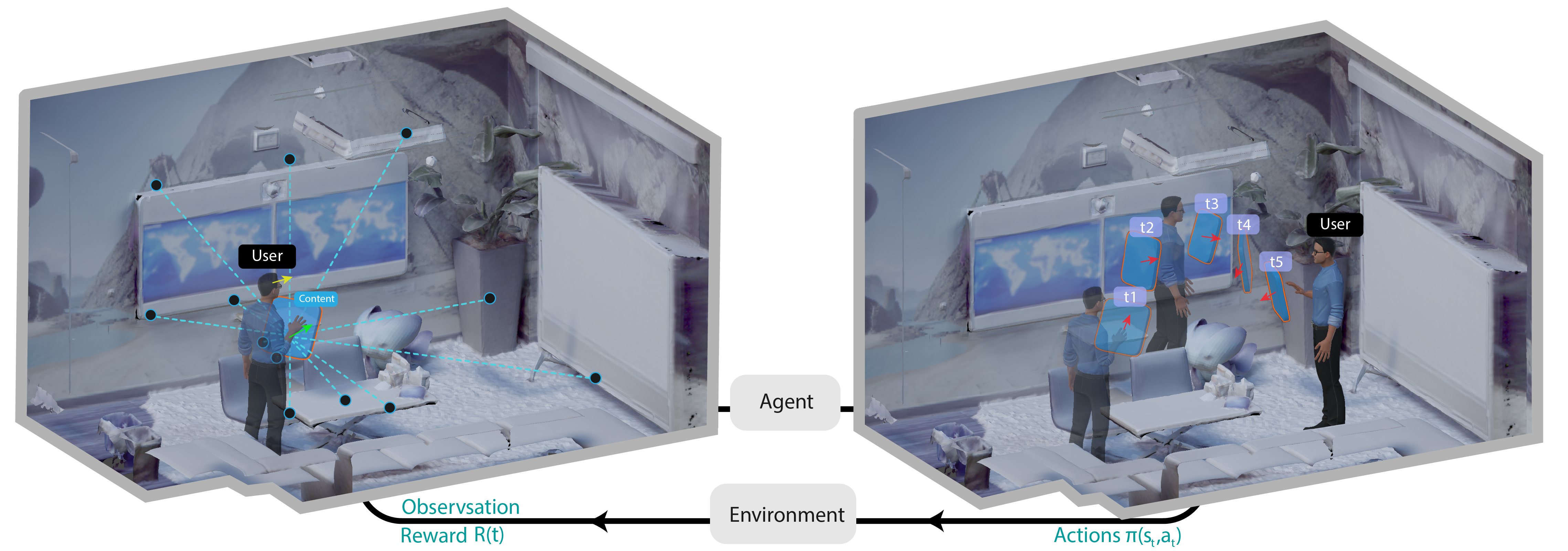}
  \caption{A demonstration of our reinforcement learning (RL) framework to assist 3D UI placement in dynamic  mixed reality environments. According to a RL agent's observations of user and environmental states (i.e., distance to surrounding environment, the user, and the user's pose), the RL agent generates a force vector (see red arrow) to maneuver the content in 3D via a physics simulation, with the goal of maximizing the accumulated reward for users in mobile scenarios.}
  \label{fig:teaser}
  \Description[Reinforcement learning framework]{Our framework of using reinforcement learning to place content in mixed reality, in which an agent observes user and environmental states to make decisions about content positions in real time.}
\end{teaserfigure}


\maketitle

%% Sections for the paper
\section{Introduction}

% why MR and how MR helps
Mixed Reality (MR) technologies have the potential to assist users' tasks by integrating digital content pervasively with the users' view of their physical environment \cite{grubert17}. Users are empowered to rely on the digital information continuously on-the-go in a variety of everyday tasks. However, determining the optimal placement of a 3D user interface (UI) in the physical space poses a non-trivial challenge due to the increased degree of freedom and dynamic nature of MR use cases. Since virtual information could be displayed anywhere and anytime in 3D space, it may hinder real-world activities by demanding unnecessary user attention and manual interactions (e.g., moving or hiding digital content) if not designed carefully and updated continuously. Such issues become prominent especially when users are mobile, as an initially useful placement may lose its utility after users alter their poses or spatial positions. In order for MR UIs to better support the users, there are two unsolved challenges: (1) they must seamlessly adapt to dynamic contextual changes such as user's position, pose, and surroundings, which traditional rule-based adaptations fall short of doing \cite{lindlbauer22}; and (2) they must take multiple adaptation goals into consideration in real time, such as visibility, reachability, and comfort \cite{belo2022,john2023}.
% \hh{The last sentence confuses me. How does these two examples (productivity and casual info acquisition) related to object placement?} \fl{Good point, I made better connections and bridged two paragraphs}

% \hh{Object placement has two keys problems. First, dynamic changing of the environment make it hard to apply rule-based placement. Second, special consideration for users' experience such as comfort, easy to interact, etc. Could we include refs saying that these two problems are indeed challenges? maybe merge this paragraph with the next.} \fl{I like it, I included them at the end of last paragraph}
% \hh{limitations of optimization-based method and how and why RL could possibly overcome them.}
% \hh{optimization-based method are mostly used for discretized space.}

% recent approaches to address this and limitations
To realize such adaptive MR UI behaviors, recent research formulates the problem as multi-objective optimization \cite{cheng21,han23,lindlbauer19,cheng23,belo2022,belo21,johnea2023,john2023}. User's goals are formulated as a set of objective functions and placements are selected that maximize/minimize these objectives. However, there are two theoretical challenges with optimization-based solutions: (1) the optimization needs to be re-solved each time users move themselves in space, which may be computationally costly and inefficient; (2) these solutions may fail to generalize to other environments or user movement patterns, requiring either reconfiguration or modifications of objectives and constraints. 

% \hh{non-linear optimization does not necessarily associated with real-time solution. How does a genetic algorithm give real-time solutions?}
% \hh{maybe we should move this to related work.}
% \hh{I think we should emphasize two challenges on users' behaviors: constant movements and generalizability to new movement/ environment. For both case, existing optimization algorithms have to be re-solved every time. The fundamental difference is that, existing optimization give a solution of the alignment given users' position and current environment; however, our RL gives a function that could determine the alignment given users' positions and the environment. }

%However, it may take a long time to search for a near-optimal solution especially in dynamic and complex environments \cite{amine19}. In short, current solutions fall short to (1) derive optimal placements of MR content continuously in real-time; (2) take into account the utility of the content temporally downstream, when users constantly move around in the physical space, introducing dynamic changes in the their surrounding environments.

% our approach and its novelty
In this work, we take a novel perspective to address the 3DUI placement challenge in MR. Inspired by its successful applications in dynamic real-world environment such as robotics and automated vehicles, deep reinforcement learning (RL) is employed to make decisions about placement of 3D UIs in dynamic scenarios. As compared to optimization-based approaches in which objectives are defined and observed more intuitively, RL demands careful crafting of the reward functions and a large amount of trials and errors during training. In contrast, it could be more advantageous in handling complex and dynamic scenarios with opportunities to incorporate user preferences, especially when decisions need to be made sequentially. 
%Unlike optimization-based methods that relies on alignment predefined by objective functions, RL learns a policy that symbolizes such alignment and could be more flexible to handle dynamic use cases. 
By observing the environments and user states, our proposed RL agent learns to take actions, observe the outcome, and formulate a policy that maximizes long-term reward for the users through its own interactive experiences. To our knowledge, this work is the first that employed RL to place virtual information in 3D MR space when users are mobile. We describe in detail the training configuration and the setup of the models. To assess the potential of our RL-driven UI placement approach in MR, we validate the performance and generalizability of our trained model with a preliminary evaluation. Our results demonstrate the potential of RL to address the 3D UI placement challenges in MR environments by helping users decide optimal placements that maximize the accumulative reward on-the-go. Additionally, we highlight future possibilities about how to further improve our proposed method with state-of-the-art RL methodologies.

% emphasize the contributions
In summary, the contributions of our work are three-fold: (1) a novel approach proposed to adapt MR UI placement through state-of-the-art RL algorithms; (2) a preliminary simulation-based evaluation of our proposed method; (3) future directions identified that leverage RL for personalized and adaptive UI placement in MR.

% Notes before writing the full intro  
% Reinforcement learning is an important paradigm in machine learning in which computing agents learn to make optimal decisions by interacting with an environment, conducting actions, and receiving appropriate rewards. Mixed Reality blends the physical and digital spaces together, encompassing both virtual reality and augmented reality. Content layout in XR refers how to position and scale virtual elements in VR and AR in different environments such as rooms, open spaces et cetera. These virtual elements may be any objects in AR or VR space. Content Layout optimization involves placing human agents optimally such that they can easily communicate/see with other human agents and interact with their environment efficiently.

% Previously, RL has been used successfully to solve several problems in the XR domain, such as agent navigation and interaction, dynamic content generation, robotics training, social interaction enhancement etc. (note: insert all citations).

% Paragraph 3: Model-free and "with model" rl algorithms, types of model-free algorithms, policy optimization and value optimization, q learning as an advanced form of value optimization. Policy optimization ---> MDP (Markov Decision Process) modeling; optimizing long term reward etc. Problems with reinforcement learning. Discrete actions vs continuous actions.

% Paragraph 4: briefly - describe the problem , how it is solved, PPO and A2C, facebook reality labs dataset, results

\section{Related Work}

% \subsection{Information displays in mixed reality}
% %highlight challenges on-the-go when user is mobile
% As compared to traditional 2D screens, AR/MR opens more possibilities for information displays due to its higher dimensionality and rich interaction paradigms. Existing strategies can be categorized into the following categories: world-fixed, object-fixed, head-fixed, body-fixed, or device-fixed \cite{laviola20173d,feiner93}. World-fixed frame of reference are commonly used in AR/MR interfaces, though they limit the accessibility of the virtual information when users are mobile. Motivated by this, research has been investigating mobile-friendly user interfaces. Ens et al. proposed a mobile window management system that adapts its reference frames across body, world, and hand \cite{ens14}. Lages \& Bowman proposed adapting the reference frame of virtual windows through controller input \cite{lages19}. Lu et al. proposed head and body-referenced peripheral information displays \cite{lu20,lu21}. However, these solutions are manual and static by maintaining a fixed configuration unless instructed by the users, making them more \textit{adaptable} than \textit{adaptive} \cite{findlater04}. Though an \textit{adaptable} interface offers more user agency, it requires users to make decisions about when and how to adapt, which could be undesirable in MR use cases. In this work, we focus primarily on \textit{adaptive} interface solutions that act towards contextual changes spontaneously by observing users and their environments in real time.

\subsection{Adaptive user interfaces in mixed reality}
%mention existing approaches for doing so, such as MOO linear/nonlinear/combinatorial optimizations, lack of training dataset
%view management
As compared to traditional 2D screens, AR/MR opens more possibilities for information displays due to its higher dimensionality and rich interaction paradigms. View management refers to how virtual information should be displayed and managed in the user's viewport \cite{bell01}. Research in view management has primarily investigated label placement, which investigates how 2D UIs should be arranged in the screen plane \cite{azuma03,orlosky15,koppel21,bekos2019,tatzgern14}, with less emphasis on content registered in the 3D environment. Grubert and colleagues proposed the concept ``Pervasive AR'', in which they argue that future AR/MR experiences should be context-aware to continuously adapt to the users' environments and tasks \cite{grubert17}. However, it has been challenging to develop such interfaces due to the dynamic nature of MR use cases, the frequent conflicts in user goals, and the theoretically infinite possible placements in 3D space.

%gneralizability and continuously change

MRTK\footnote{https://learn.microsoft.com/en-us/windows/mixed-reality/mrtk-unity/mrtk2/} provides solver components to decide placement of UIs based on one or more objectives. However, the adaptations are mostly rule-based executions. The objectives are solved only sequentially which makes it less applicable in dynamic multi-objective scenarios. Recent work proposes to realize such adaptive behaviors for 3D content in MR through optimization, which have been widely studied in 2D applications to decide the layout of UI elements \cite{dayama20,todi16,zhai02,oulasvirta20}. For example, Lindlbauer et al. implemented an adaptive UI system in MR using integer programming taking into account the user's task and cognitive load \cite{lindlbauer19}. Belo et al. explored placement of UIs in MR that maximizes ergonomic comfort for users \cite{belo21}. Cheng et al. explored automatic placement of MR content while optimizing visibility, consistency, and semantic correlations using linear programming \cite{cheng21}. Similar methods were applied in a recent work by Han et al. to blend MR windows on top of physical object meshes, taking into account the properties of physical objects \cite{han23}. However, adapations in these work mostly took place in a discrete solution space, which may not guarantee optimality when users are moving continuously. To enable adaptations in continuous spaces, Belo et al. explored simulated annealing for solving non-linear objectives for MR UI adaptations. Johns et al. explored genetic algorithms for deriving optimal placements along the pareto frontier \cite{john2023,johnea2023}. This work demonstrated the potential of optimization-based approach for UI adaptations in MR.

In this work, we take a new perspective by applying RL to assist with content placement in MR environments, by training an AI agent how to ``drive'' a piece of UI content to maximize user reward.

\subsection{Reinforcement learning for interface adaptations}
%existing applications of RL in games, simulation, robotics, and why it could be a new perspective for content placement in MR
RL is a category of machine learning that enables intelligent agents to learn how to take actions in a dynamic environment with the goal to maximize long-term rewards temporally downstream. It has demonstrated great potential in scenarios where decisions need to be planned sequentially, such as robotics \cite{jens2013}, automotive vehicles \cite{zhang18,fayjie18}, gaming \cite{kaiser19}, and crowd simulations \cite{charalambous23}. We argue that RL could be a suitable approach to address the 3D UI placement challenge in MR due to its dynamic and user-centric nature. Unlike supervised learning which relies on large amount of labelled data, RL learns by observing and interacting with the environment on itself, which makes it powerful (1) for handling complex, unexpected, and dynamic scenarios, (2) when current decisions can influence future states, and (3) when large-scale labelled training datasets are challenging to obtain. Due to such properties, RL has been recently explored as a new strategy to achieve interface adaptations in both 2D and MR spaces.  Todi et al. used model-based RL to adapt item ranking in a 2D menu \cite{todi21}. Yu et al. studied using RL to determine for timing of intelligent assistance during object selection tasks in MR \cite{yu23}. Gebhardt et al. explored using RL to adapt the visibility of 2D labels in VR based on gaze data \cite{gebhardt19}. Chen et al. employed RL for AR label placement in the view plane during basketball matches \cite{chen23}, which we consider the most relevant to our work. Their results show promising potential of RL to intelligently assist 2D UI placements in 3D environments when the user's viewport is static. However, there has been a lack of exploration of how RL could be applied to assist 3D UI placement and adaptation, where (1) UIs could be registered anywhere in the 3D physical world beyond a 2D camera plane and (2) the user's egocentric viewport dynamically changes in mobile scenarios. Our work sheds light on such possibilities by training a RL agent that is capable of maneuvering UIs in 3D constantly according to its observations of how the users' and environmental states change.

\section{Problem Formulation}

%notation. 
% \subsection{Partially-Observable MDP}
% model-based vs model-free
% Shawn
% \hh{Why we can frame this problem into a RL problem. The issue there is that our current problem is not a sequential problem. }

% \hh{human move => input perturbation. 
% 1. completely re-solve the optimization problem.
% 2. how input perturbation influence the solution. }
% \hh{any paper comparing simulated annealing vs. genetic algorithm vs. RL?}
% \hh{limitations of optimization-based method and how and why RL could possibly overcome them.}
% \hh{optimization-based method are mostly used for discretized space.}

\subsection{Environment \& User Simulation}
% Mengyu & Feiyu
To construct a realistic training environment for the RL agent to freely explore and learn, we utilize the Replica dataset \cite{straub2019replica}, which contains high-fidelity 3D scans of indoor scenes with segmented objects and bounding boxes. This grants us full understanding of a simulated physical environment, including position, orientation, and size of floors, walls, and furniture. We picked four scenes from the dataset covering small/large office/living room spaces, two for training and two for validation. In our simulated environment, we replicate the actions of a simulated user traversing the space thoroughly and adequately. The simulated user would make stops at random locations on a 3D Cartesian grid situated inside the room, mimicking a variety of user poses and locations. These include, but and not limited to, sitting on a sofa, standing in front of a whiteboard, sitting around a table, all while facing a wall, furniture, or empty space (see \autoref{fig:1} (a)). The RL agent is then designed to learn 3D UI adaptations that maximize the reward for the presented simulated user along the horizon. 

We understand that such approaches may introduce limitations depending on how well the simulated user behaves in comparison with actual users. We argue that our approach could bring unique benefits since real-world user trajectory data is largely unavailable in well-labeled indoor spaces. Applying synthetic user data could be a scalable strategy to train and gauge the feasibility of RL-based methods. As an initial exploration, we started with one piece of virtual content in the environment, the movement of which the RL agent takes full control of.

\subsection{RL Agent}
% actor-critic network + PPO
We employ model-free RL which does not require prior knowledge of the transition model. The RL agent leverages proximal policy optimization (PPO), a state-of-the-art deep policy gradient algorithm which has demonstrated great potential in a variety of benchmarks and applications \cite{berner2019dota,mirhoseini2020chip,schulman2017proximal}. In the following sections, we highlight our designs of three core components: (1) observations (what information the agent collects from the environment), (2) actions (what actions the agent produces to control the placement of 3D UI in MR), and (3) reward (how the agent is rewarded or penalized based on its action).

\subsubsection{Observations}
% ray sensor, distance to surroundings, user, facing direction, velocity
In each time step $t$, the agent receives the following observations: (1) $\{Ray_1, Ray_2 ... Ray_{22}\}$: twenty-two rays evenly cast around the virtual content to collect information about its surrounding in the current time step, including whether any ray hits the physical environment mesh, and if so, the start/end point of each ray hit, the hit object type, and the length of the ray. As such, the RL agent observes the content's relative position to its surrounding environment; (2) $\{Pos_c, Rot_c, Dir_c\} \in \mathbb{R}^3$: the virtual content's position, rotation, and facing direction in the local space of the environment; and (3) $\{V_c, Dist\} \in \mathbb{R}$: the virtual content's local velocity and its euclidean distance in relation to the user; (4) $\{Pos_u, Dir_u\} \in \mathbb{R}^3$: the user's current position and facing direction in space  (see \autoref{fig:teaser} left). The observations were normalized before feeding into the network.

\subsubsection{Action Space}
% 3D force vector
Based on the observations, the RL agent produces a three-dimensional continuous action $\{f_x, f_y, f_z\} \in [-1, 1]$, symbolizing a 3D force vector to apply to the virtual UI content (see \autoref{fig:teaser} right). With physic simulation, the virtual content could be freely maneuvered, accelerated, and decelerated in 3D based on the applied force vector. The RL agent only controls the content's position in space. The content was always oriented in a billboard fashion to the users, ensuring maximum visibility. The action was produced every five time steps.

\subsubsection{Reward Design}
% Mengyu & Feiyu
% algorithm box
Design of the reward functions plays critical roles in enabling the RL agent to actively query to evaluate the quality of a given action or state. In line with previous work on MR layout optimizations \cite{cheng21,lindlbauer19, belo2022,johnea2023}, we aim for 3D UI placements that are continuously visible, reachable, demonstrate realistic physical behaviors, and stable. We designed the final reward to be the aggregation of four rewards \{$R_{\text{visibility}}, R_{\text{reachability}}, R_{\text{physicality}}, R_{\text{stability}}$\}:

\begin{align}
    R_{\text{visibility}} &=
    \begin{cases}
      0.1 \cdot P_{percent}, & \text{if } P_{percent} > 0\\
      -0.1, & \text{otherwise}\\
    \end{cases}\\
    R_{\text{reachability}} &=
    \begin{cases}
      0.1 \cdot e ^ {\frac{(Dist - 0.5) ^ 4}{0.05}} , & \text{if UI is in front of the user}\\
      -0.1, & \text{otherwise}\\
    \end{cases}\\
    R_{\text{physicality}} &=
    \begin{cases}
      -0.01, & \text{if UI overlaps with physical objects}\\
      0.01, & \text{otherwise}\\
    \end{cases}\\
    R_{\text{stability}} &=
    \begin{cases}
      0, & \text{if the user is moving}\\
      0.01, & \text{if the user is stationary and  } V_{c}< \text{0.3}\\
      -0.01, & \text{if the user is stationary and  } V_{c}\geq \text{0.3}\\
    \end{cases}
\end{align}

\begin{figure*}[t]
\includegraphics[width=\textwidth]{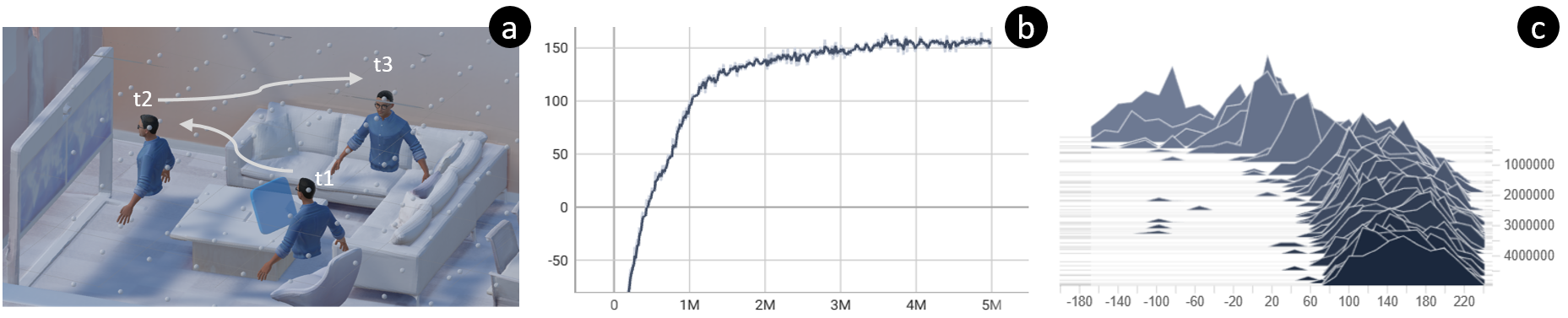}
  \abovecaptionskip=2pt
\caption{(a) Simulating user movement during training by dynamically interpolating an avatar along a 3D Cartesian grid facing a random direction every few seconds; (b) the accumulative reward per episode (y-axis) in relation to the number of total steps (x-axis); (c) the histogram of the reward distribution (x-axis) in relation to the total number of steps (y-axis).}
\label{fig:1}
\Description[Training configuration and results]{Left: our training configurations with a simulated user randomly traversing a 3D grid; middle: a graph showing the average accumulated reward in relation to the number of steps. The reward increases and stabilizes near the end of the training; right: a graph representing the histogram of reward in relation to the number of steps. The reward gradually converges towards the end.}
\end{figure*}

We acknowledge that designing quality reward functions which empower the agents to learn, explore, and exploit demands non-trivial effort in the domains of RL. The current reward design was developed after extensive empirical testing and benchmarks. $P_{percent}$ represents the aggregation of percentages of the pixels that belong to the content within the user's camera render texture and the proportion of the content that is visible for the current time step (i.e., how much of the content is visible and how much it occupies the user's viewport); $Dist$ represents the current euclidean distance from the content to the user; $V_c$ represent the current velocity of the content measured by meters/second. The reachability reward applied a kernel function, as inspired by \cite{cheng21}. The scale of the rewards (i.e., 0.1, 0.01) followed prior guidelines \cite{chen23,gebhardt19,chen19} and our testing. The thresholds for $V_c$ and $P_{percent}$ were determined empirically.

\subsection{Training Configuration \& Results}
% episode ending criteria. random initialization
% Shawn

The Unity3D engine was employed to simulate the 3D training environment, which consists of the simulated physical environment, the user, and the virtual content. The ML-Agents toolkit\footnote{https://github.com/Unity-Technologies/ml-agents} was leveraged for training the RL agent \cite{juliani2020}. The toolkit bridges information collected in the Unity3D engine with a PyTorch backend to run the deep RL training.

The PPO training network consists of (1) an encoder which encodes the observations and states of the simulation environment; (2) an actor network to learn the optimal policy $\pi(s_t, a_t)$ for the agent, and (3) a critic network which estimates the accumulated reward $R(t)$ for a given action. Both the actor and critic networks were designed with two fully-connected hidden layers with 128 units. The training was configured with six parallel training instances to speed up the process. Each episode consisted of training for around one thousand steps, concluding once the maximum step limit was reached. The entire training lasted for five million steps in total, taking around six hours on a desktop PC with a dedicated Nvidia GeForce RTX3080 GPU, an Intel i9 24-core CPU, and a 64 GB RAM. The learning rate was initialized as $3e-4$, which decayed linearly until max step is reached so the training converges stably. We used a buffer size of 409,600 and a batch size of 2,048 for the agents to gather enough information from its experiences to learn the policy. \autoref{fig:1} (b) shows the average accumulated reward per episode in relation to the number of total steps, and \autoref{fig:1} (c) shows the histogram distribution of the rewards. The reward increased effectively during training while its standard deviation decreased, indicating the successful learning of a policy by the RL agent.

% \subsection{Results}
% Learning curriculum
% highlight unique benefits such as generalizability to environmental changes

\section{Preliminary Evaluation}
% collect number of collisoins, visible spots, time taken to go to new space etc. measures
% generalize to different environments
% dynamic environment
\begin{figure*}[t]
\includegraphics[width=\textwidth]{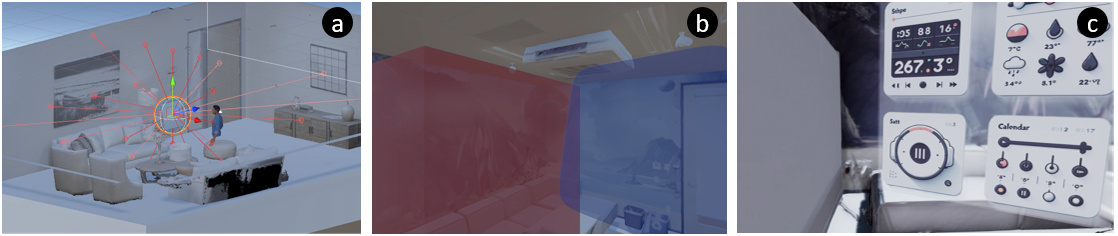}
   \abovecaptionskip=2pt
 \belowcaptionskip=-10pt
\caption{(a) A snapshot of the training in which a UI driven by RL gauges its distances to surroundings and simulated user poses; (b) a snapshot of the dynamic obstacle test, in which the content (blue square) maintains visibility, reachability and avoiding collision with a presence of a randomly moving obstacle (red cube); (c) a snapshot of the placement result running our trained RL model.}
\label{fig:2}
\Description[Training snapshot and validation demo]{Left: A snapshot of the training scene in which the content shoots out rays to detect the scene and the user to place itself; middle: a snapshot of the model running in real time, in which a reinforcement learning agent controls the UI to avoid it overlapping with a moving obstacle; right: a snapshot of a piece of UI being placed in real time by our RL model, in which a UI content is automatically placed next to a whiteboard in user's view while avoiding collisions.}
\end{figure*}

\subsection{Setup}
We conducted an initial evaluation on our model by placing the RL agent in each of the four Replica environments (two training and two validation environments) and running a simulation continuously for ten thousand steps, with a simulated virtual user randomly traversing the environment. The RL model was evaluated based on three considerations:
\begin{itemize}[leftmargin=*, noitemsep, topsep=0pt]
    \item \textbf{\textit{Heuristics}}: A set of heuristics to evaluate the placement continuously derived by the RL model, including visible UI \% (i.e., average percent of the UI that is visible to the user each time step), non-collision \% (i.e., percentage of steps that the UI does not overlap / collide with any physical object), distance offset  (i.e., average absolute distance offset from the UI to the user in meters as compared to a half-meter baseline), and speed (i.e., the average speed of the content per time step in meters per second).
    \item \textbf{\textit{Generalizability}}: In addition to the training scenes, we picked two additional indoor scenes as validation sets. We collected the same heuristics for the different indoor scenes to see if the agent could maintain ideal results while experiencing an environment that it has no experience in.
    \item \textbf{\textit{Dynamics}}: We introduced moving obstacles to symbolize moving  people or objects that the RL-controlled UI should avoid. The obstacle was represented by a 1 by 1 by 1 meter cube which moved randomly in the environment (see \autoref{fig:2} (b)). We collected the same heuristics with and without the presence of these dynamic obstacles to gauge the ability of the model to react to unexpected dynamic changes that take place around the users. 
\end{itemize}

\begin{table*}
\caption{The statistics showing the average value of each heuristic in the simulation testing (T for training, and V for validation).}
% \vspace{-5pt}
\begin{tabular}{ c | c c | c c | c c | c c } 
 \hline
   & \multicolumn{4}{c|}{Static Environments} & \multicolumn{4}{c}{Dynamic Obstacles} \\ 
 \hline
   & Env1 (T) & Env2 (T) & Env3 (V) & Env4 (V) &  Env1 (T) & Env2 (T) & Env3 (V) & Env4 (V) \\ 
   \hline
 Visible UI \% & 93.28\% & 91.74\% & 86.50\% & 84.05\% & 85.27\% & 88.05\% & 83.38\% & 82.92\%\\ 
 \hline
 Non-Collision \% & 94.09\%  & 94.80 \% & 87.06\% & 82.18\% & 82.37\% & 90.32\% & 84.48\%  & 85.86\% \\ 
  \hline
 Distance Offset & 0.06 & 0.04 & 0.11 & 0.21 & 0.11 & 0.17 &  0.22 & 0.31\\ 
 \hline
  Speed & 0.68  & 0.72 & 0.69 & 0.74 & 0.72 & 0.75 & 0.77 &  0.84 \\ 
 \hline
\end{tabular}
\label{tab:1}
\end{table*}
\subsection{Results}
As shown in \autoref{tab:1}, our results demonstrated the potential of RL-based approach for adaptive 3D UI placement. In more static environments, for the environments that the agent was trained upon, more than 91\% of the UI was continuously visible. It did not collide with any environmental mesh more than 94\% of the time. It also maintained an arm-reachable distance to the users. In the two validation environments that the agents have not experienced, our results show that UI could still stay in reasonable locations with around 85\% of itself being continuously visible to the users each time step. It maintained a distance slightly further away from the user, but still within a reachable distance. Similarly, in dynamic settings when moving obstacles are introduced in the environments, our model still achieved good placements in both the training and validation environments with negligible performance drops especially on the validation environments, which the drops of visibility/physicality were capped at 3.12\%. The UI could maintain its reachability level even when the obstacle slides in view (see \autoref{fig:2} (b)), and avoid most upcoming collisions to keep its visibility to the users. As a result, the average distance from the UI to the user slightly increased to compensate for such trade-offs. This demonstrates the RL agent's ability to adapt and generalize to environments and certain degrees of unexpected changes that it has not experienced before. However, we are aware that the higher performance drops on the two training environments with dynamic obstacles might be a sign of overfitting. Future research is required to further validate our model in more diverse environments. \autoref{fig:2} (c) showcases the output of the model in real time, in which RL places the UI while avoiding collision with the whiteboard, in the mean time ensuring visibility and reachability to the users.

One limitation we observed is that the UI still moved at a relatively high speed, with an average more than the level we set in the reward function. One reason for this could be that our simulation symbolized a busy setting in which users relocate frequently, so there were less opportunities for the RL agent to experience the static reward. Future work is needed in terms of how to further increase the stability of the UI. 

\section{Future Opportunities}
% More than 1 content with MA-POCA
% A posteriori approach
% RLHF
% Comparison with extsing opmization methods
% model -based vs model-free, with HCI models

We are excited about the positive outcome of our initial exploration. Here, we would like to highlight future opportunities and challenges for leveraging RL for content placement in dynamic 3D MR environments:

\begin{itemize}[leftmargin=*, noitemsep, topsep=0pt]
    \item \textit{More considerations on reward formulations.} As an initial exploration, our current RL setup did not consider other aspects while formulating the reward, such as semantics \cite{cheng21} and affordances  \cite{xiao18,cheng23} of physical objects, content type (e.g., 2D vs 3D, text-heavy vs. images), spatial consistency \cite{barretspatial}, and user ergonomics \cite{belo21}. Recent work has highlighted the benefits of such considerations. Future research should take into account these aspects and explore how to incorporate them into current reward designs, as well as how to further improve stability of the UI content. 
    
    \item \textit{VR simulation vs. actual AR/MR settings.} Our training assumes good understanding of environmental and user status by simulating AR/MR environments in VR. Training in actual AR/MR spaces may surface more challenges due to the extensive trials and errors RL needs and the imperfect understandings of the physical world. Future work is needed in order to bridge the gaps between simulated and actual user behaviors. For example,  utilizing generative AI models, synthetic data could be generated from user's actual movement trajectories for training the RL agent \cite{chu23}. As such, the agent could learn from common user behavioral patterns to further optimize downstream performance.
    
    \item \textit{Multiple users and UIs.} Our exploration demonstrated the initial feasibility of RL controlling the placements of one piece of 3D UI content. However, users frequently need access to multiple applications with different relevancy to their tasks. To support multiple content elements, multi-agent RL frameworks could be leveraged for training cooperative behaviors among multiple agents so they make the best trade-offs \cite{lowe2017multi,cohen2021use}. Same strategies may also apply to multi-user collaborative settings. Future research could investigate the feasibility of these directions.
    
    \item \textit{RL with human feedback (RLHF) for preference learning.} Research has demonstrated that AI models with better performance may not be considered more useful by the users \cite{roy19}. Similarly, the quality of 3D UI placements in MR from the user's perspective may often be highly subjective. Thus, it is critical to derive placements that align well with user expectations, preferences and agency \cite{lu22transition}. Recent work has revealed the potential of RL to incorporate human-in-the-loop preference learning with a handful of queries, which generalizes well to new tasks without having to retrain the models \cite{christiano2017deep,hejna2023few}. This empowers preference elicitations on-the-go, which could be more flexible as compared to constraints-based methods in optimizations. Future research could further investigate the potential of RLHF for deriving personalized and preferable UI placements in MR spaces.

    \item \textit{Model-based RL.} Our exploration leveraged model-free RL, which could be a powerful technique for tackling real-world situations when the environment is noisy, complex, and unpredictable. The downside is model-free RL requires an extensive amount of trials for the agent to learn a good policy. In the field of HCI, a wide variety of models have been developed to predict human motor and cognitive performances \cite{fitts1954information,hick1952rate,florent21}. This makes \textit{model-based RL} viable as an alternative, in which the RL agent simulates consequences of its actions without executing them to plan ahead accordingly. As such, the training could be done more efficiently. Recent work by Todi et al. demonstrated the feasibility of \textit{model-based RL} for adapting 2D menus leveraging models such as Fitts' law \cite{todi21}. Future research could further explore such directions for MR use cases, in which user performance models are leveraged.
    
    \item \textit{Comparisons with optimization-based methods.} Due to the scope of this work, we did not compare RL with optimization-based methods which demonstrated great potential in recent MR adaptive UI work, such as linear/integer programming \cite{cheng21,han23,lindlbauer19}, simulated annealing \cite{belo2022}, and genetic algorithms \cite{john2023,johnea2023}. Future research is needed to explore the trade-offs among these method in both experimentally-controlled and ecologically-valid settings of MR UI scenarios.
    
    \item \textit{More explorations on generalizability.} Though our results provide initial evidence on the generalizability of our RL-based approach, the training was conducted in less diverse environments. Introducing a higher degree of variability and intricacy may potentially facilitate a more robust learning process (e.g., multi-floor environments with moving bystanders ), allowing the model to adapt more effectively to diverse UI scenarios. We aim to simulate more accurate real-world scenarios, thereby enabling the model to acquire a broader range of behaviors.
\end{itemize}

\section{Conclusion}
In this work, we explored the potential of reinforcement learning (RL) for 3D user interface (UI) placement in mixed reality (MR) environments. By having a RL agent interactively engaging with a simulated physical environment with trials and errors, it could learn a policy that assist the placements of 3D UIs in order to maximize the utility for the users. We trained a RL agent that is capable of assisting the placements of 3D UI continuously, demonstrated the potential of our approach through a preliminary simulated evaluations that explore the generalizability of the model to new and dynamic environments, and called attention to challenges such as UI stability and overfitting. Based on our results, we further highlight opportunities for future research that explore RL for personalized and adaptive information displays in MR environments.

% \begin{acks}
% \end{acks}

\bibliographystyle{ACM-Reference-Format}
\bibliography{sample-base}

\section{Appendix - Disclaimer}
This paper was prepared for informational purposes by the Global Technology Applied Research center of JPMorgan Chase \& Co. This paper is not a product of the Research Department of JPMorgan Chase \& Co. or its affiliates. Neither JPMorgan Chase \& Co. nor any of its affiliates makes any explicit or implied representation or warranty and none of them accept any liability in connection with this paper, including, without limitation, with respect to the completeness, accuracy, or reliability of the information contained herein and the potential legal, compliance, tax, or accounting effects thereof. This document is not intended as investment research or investment advice, or as a recommendation, offer, or solicitation for the purchase or sale of any security, financial instrument, financial product or service, or to be used in any way for evaluating the merits of participating in any transaction.

\end{document}